# Electrically tunable multi-terminal SQUID-on-tip

*Aviram Uri[1]\*, Alexander Y. Meltzer[1], Yonathan Anahory[1], Lior Embon[1,†], Ella O. Lachman[1], Dorri Halbertal[1], Naren HR[1], Yuri Myasoedov[1], Martin E. Huber[1,2], Andrea Young[1,3], and Eli Zeldov[1]\**

[1]Department of Condensed Matter Physics, Weizmann Institute of Science, Rehovot 7610001, Israel

[2]Departments of Physics and Electrical Engineering, University of Colorado Denver, Denver, Colorado 80217, USA

[3]Department of Physics, Broida Hall, University of California, Santa Barbara, CA 93106-9530

**ABSTRACT**

We present a new nanoscale superconducting quantum interference device (SQUID) whose interference pattern can be shifted electrically *in-situ*. The device consists of a nanoscale four-terminal/four-junction SQUID fabricated at the apex of a sharp pipette using a self-aligned three-step deposition of Pb. In contrast to conventional two-terminal/two-junction SQUIDs that display optimal sensitivity when flux biased to about a quarter of the flux quantum, the additional terminals and junctions allow optimal sensitivity at arbitrary applied flux, thus eliminating the magnetic field "blind spots". We demonstrate spin sensitivity of 5 to 8 $\mu_B/Hz^{1/2}$ over a continuous field range of 0 to 0.5 T, with promising applications for nanoscale scanning magnetic imaging.

In recent years, there has been a growing effort to develop and apply nanoscale magnetic imaging tools in order to address the rapidly evolving fields of nanomagnetism and spintronics. These include magnetic force microscopy (MFM)[1,2], magnetic resonance force microscopy (MRFM)[3–5], nitrogen vacancy (NV) centers sensors[6–9], scanning Hall probe microscopy (SHPM)[10–12], x-ray magnetic microscopy (XRM)[13], and micro- or nano-superconducting quantum interference device (SQUID)[14–20] based scanning microscopy (SSM)[21–32]. Scanning micro- and nanoscale SQUIDs are of particular interest for magnetic imaging due to their high sensitivity and large bandwidth[15,19]. The two main technological approaches to the fabrication of scanning SQUIDs are based on planar lithographic methods[21,26,33–36] and on self-aligned SQUID-on-tip (SOT) deposition[22,24,37].

In the planar SQUID architecture, spatial resolution is limited but pickup and modulation coils can be integrated to allow operation of the SQUID at optimal flux bias conditions using a flux-locked loop (FLL) feedback mechanism[15,18,19,21,33,38,39]. Because the magnetic field of the sample is not coupled to the SQUID loop directly, but rather through a pickup coil, integration of a modulation coil or an integrated current-carrying element[15,19,21,33,38,39] allows the total flux in the SQUID loop to be maintained at its optimal bias while the magnetic field of the sample is varied independently.

SOTs, in contrast, have better spatial resolution due to their small size and close proximity to the sample, attain higher spin sensitivity, and can operate at high magnetic fields[24]. The nanoscale proximity of the SOT to the sample surface, which is its key advantage, dictates however that the flux in the SQUID loop is directly coupled to the local field of the sample and therefore cannot be modified independently. As a result, the FLL concept cannot be implemented in direct nanoscale magnetic imaging. This poses a significant drawback, since the high sensitivity of the SOT is achieved only at specific field values–leaving substantial "blind spots" at intermediate magnetic fields.

We present here a new approach that heals these blind spots by current-controlled phase bias of a multi-terminal/multi-junction SQUID. Instead of controlling the magnetic flux in the SQUID loop, we use a multi-terminal/multi-junction SQUID-on-tip (mSOT) in which we can electrically control in-situ the phase of the superconducting order parameter across the junctions. The resulting phase control provides a highly sensitive SQUID operation at optimal bias conditions at any value of the applied field and opens the possibility of implementing a superconducting-phase locked loop (SPLL), the phase bias equivalent of the FLL commonly used with two-terminal/two-junction SQUIDs.

The operation of the mSOT is based on the incorporation of extra degrees of freedom by integrating additional junctions and terminals in the SQUID loop as shown schematically in Fig. 1a. Multi-junction devices, as well as multi-terminal configurations, have been proposed and used in the past for logic circuits[40], enhanced SQUID capabilities[37,41,42], voltage rectification[43,44], current amplification[45], kinetic inductance modulation[46], flux pumps[47], and flux modulation[18]. Here, in contrast, the extra degrees of freedom are used for phase rather than flux control of the interference pattern of the SQUID.

The basic principle of the mSOT phase-biased operation can be understood by considering an ideal two-terminal/two-junction SQUID with no inductance, which is described by

$$\varphi_1 + \varphi_2 + 2\pi \frac{\Phi_a}{\Phi_0} = 2\pi n, \qquad (1)$$

where $\varphi_i$ is the superconducting phase difference across each of the Josephson junctions, $\Phi_a$ is the applied flux in the loop, $\Phi_0 = h/2e$ is the flux quantum, and $n$ is an integer. In the following we consider $n = 0$ for simplicity. The requirement for maximum total supercurrent through the SQUID sets $\varphi_1 - \varphi_2 = \pi$, resulting in the well-known quantum interference pattern $I_c = 2J_0|\cos \pi \Phi_a/\Phi_0|$, where $J_0$ is the maximal supercurrent each junction can support. The highest sensitivity of the SQUID is achieved[48] when $|\varphi_1 + \varphi_2| \approx \pi/2$, which corresponds to $|\Phi_a| \approx \Phi_0/4$. The two phases are therefore uniquely defined, hence operation at the optimal sensitivity can be achieved only if the FLL maintains the net flux in the SQUID loop at $|\Phi_a| \approx \Phi_0/4$.

Figure 1a presents a schematic mSOT configuration consisting of four identical junctions and four terminals. The primary bias current $I_1$ is applied to terminal 1, while terminals 2 and 4 are used to provide the superconducting phase control currents $I_2$ and $I_4$, respectively. In this structure, Eq. 1 is replaced by

$$\varphi_1 + \varphi_2 + \varphi_3 + \varphi_4 + 2\pi \frac{\Phi_a}{\Phi_0} = 0, \qquad (2)$$

where we took $n = 0$ for simplicity. In an mSOT, the current through terminal 1 is determined by only two of the four phases: $I_1 = J_1 - J_4 = J_0(\sin \varphi_1 - \sin \varphi_4)$. Therefore, a deviation of $\Phi_a$ form the optimal value $\Phi_0/4$ can be compensated by adjusting $\varphi_2$ or $\varphi_3$ according to Eq. (2) to preserve the original $I_1$. The result is a shift of the interference pattern $I_{c1}(\Phi_a)$ that allows the mSOT to maintain its optimal sensitivity despite the change in $\Phi_a$. The adjustments of $\varphi_2$ or $\varphi_3$ are implemented by using the control currents $I_2$ or $I_4$ as described below.

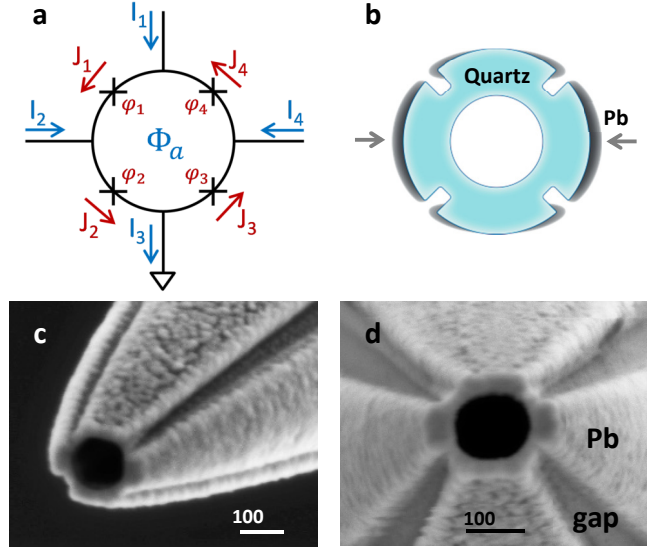

**Figure 1.** Structure of the mSOT. **(a)** Schematic diagram of an mSOT showing a SQUID loop with four junctions and four terminals. **(b)** Schematic cross section of the quartz tube with its four grooves and Pb film formed by two side depositions along the directions indicated by the arrows. **(c, d)** Scanning electron micrographs of device A showing side view **(c)** and top view **(d)** of the apex ring. Four weak superconducting links are formed by constrictions at the groove locations.

In the absence of control currents, the critical current $I_{c1}(\Phi_a)$ reaches its maximum value of $I_{c1}^{max} = 2J_0$ at zero applied flux $\Phi_a = 0$, where junctions 1 and 4 each carry their maximal dissipationless current $J_0$ inducing phase drops of $\varphi_1 = -\varphi_4 = \pi/2$. Let us for simplicity fix $I_4 = 0$, so that $J_3 = J_4$ and hence $\varphi_3 = \varphi_4$. The control current $I_2$ is then $I_2 = J_2 - J_1 = J_0(\sin\varphi_2 - \sin\varphi_1) = J_0(\sin\varphi_2 - 1)$. For any applied flux $\Phi_a$, we can keep $\varphi_1 = -\varphi_4 = -\varphi_3 = \pi/2$ fixed and modify $\varphi_2$ by adjusting $I_2$ to fulfill Eq. 2. Since $\varphi_1$ and $\varphi_4$ have not changed, the current $I_1$ of the mSOT remains at its maximal critical value; namely, $I_{c1}(\Phi_a, I_2) = I_{c1}(0,0) = I_{c1}^{max} = 2J_0$, where the control current $I_2$ required to keep $I_{c1}$ at its maximum is given by

$$I_2 = J_0(\sin\varphi_2 - 1) = J_0(\sin\left(\frac{\pi}{2} - 2\pi\frac{\Phi_a}{\Phi_0}\right) - 1) = J_0(\cos\left(2\pi\frac{\Phi_a}{\Phi_0}\right) - 1). \quad (3)$$

This result shows that the quantum interference pattern $I_{c1}(\Phi_a)$ of the mSOT can be readily shifted as a whole by the electric means of applying superconducting phase control currents $I_2$ and $I_4$. In particular, for any value of the applied flux $\Phi_a$, control currents exist that will bias the mSOT to the most sensitive working point, thus eliminating the blind spots. A more detailed theoretical study shows that this powerful control is applicable to any SQUID with at least three junctions and three terminals, including asymmetric junctions, non-sinusoidal current-phase relations, and the presence of finite inductance[49].

Note that due to its very small size, the geometric inductance of the SOT is about two orders of magnitude smaller than its kinetic inductance[22,24,49] (see S2 for details). As a result, the control currents affect the superconducting phases across the junctions with negligible change to the self-induced flux in the SQUID loop.

SOTs are fabricated by self-aligned two-sided deposition of superconducting leads along a pulled quartz pipette followed by a third deposition on the apex ring[24,50]. The main challenge in the fabrication of the mSOT is creating nanoscale multi-terminal connections to the apex. We achieve this by using a quartz tube of, initially, 1 mm outer diameter with four 0.1 x 0.15 mm$^2$ grooves equally spaced on its circumference (Fig. 1b), which maintain their relative shape upon pulling and extend all the way to the tip. The grooves provide shadowing during the two side depositions of Pb, creating gaps separating the four leads (Fig. 1c). In the third deposition on the apex, a Pb ring with four constrictions in the gap regions is formed, creating Dayem-bridge weak links (Fig. 1d) and establishing a self-aligned nanoscale four-terminal/four-junction mSOT.

Three mSOT devices (see S1 for details) were characterized at 4.2 K using the electrical circuit in Fig. 2a. The current bias $I_{B1}$ was applied to the mSOT in parallel with a shunt resistor $R_{S1}$. The resulting current $I_1$ flowing into the mSOT is measured using a SQUID series array amplifier (SSAA)[51]. The superconducting phase control currents $I_2$ and $I_4$ are provided by applying control bias currents $I_{B2}$ and $I_{B4}$. Figure 2b presents electrical characteristics $I_1(I_{B1})$ of mSOT device A showing the measured $I_1$ vs. the applied bias current $I_{B1}$ at several values of the field $B_a$, applied perpendicular to the mSOT loop plane. For $I_1 < I_{c1}(B_a)$, the mSOT is in the superconducting state and most of the applied current $I_{B1}$ flows through the device (see S3 for details). When $I_1$ reaches $I_{c1}(B_a)$ the mSOT becomes resistive and a significant part of the current is diverted to $R_{S1}$. A number of resonances are visible at higher biases.

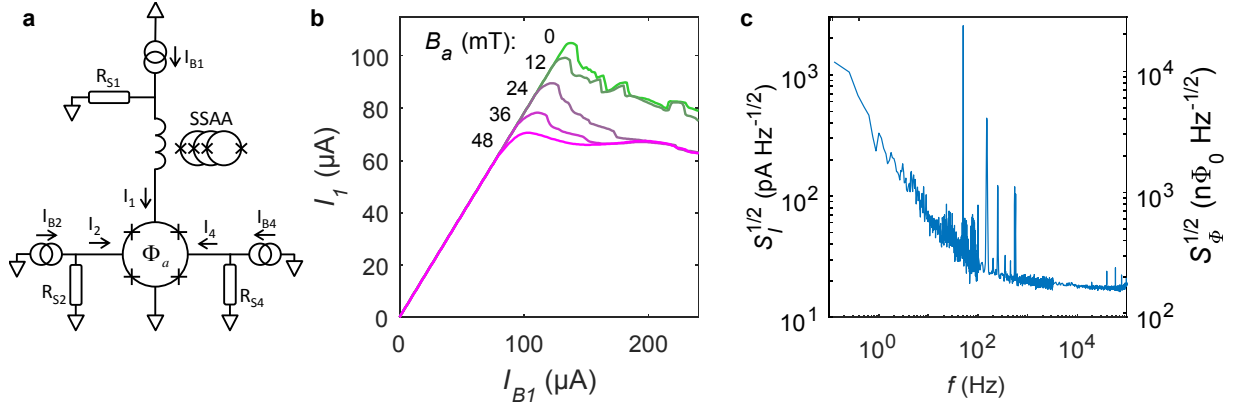

**Figure 2**. **(a)** A schematic electrical circuit of the mSOT system. **(b)** Electrical characteristics of mSOT device A at various values of $B_a$ acquired at $I_{B2} = 74\,\mu\text{A}$ and $I_{B4} = 0$. **(c)** Current noise spectral density $S_I^{1/2}$ (left axes) and corresponding flux noise $S_\Phi^{1/2}$ (right axis) at $B_a = 0.2$ T at the optimal working point.

The interference pattern of the critical current $I_{c1}(B_a)$ derived from the electrical characteristics is shown in Fig. 3a. The interference period of 97 mT, corresponding to an effective loop diameter of 165 nm, is in good agreement with the SEM image in Fig. 1d. The remarkable feature demonstrated in Fig. 3a, however, is the in-situ electric tunability of the mSOT. The three presented $I_{c1}(B_a)$ curves reveal that the entire pattern is readily shifted horizontally by varying the control bias current $I_{B2}$ (keeping $I_{B4} = 0$ for simplicity). Figure 3b demonstrates that the critical current interference pattern $I_{c1}(B_a, I_{B2})$ is shifted continuously with $I_{B2}$ and over a large span of about half a period between the lowest and highest presented values of $I_{B2}$, in agreement with the theoretical prediction[49]. This implies that, for any value of $B_a$, there is a current $I_{B2}$ for which maximum mSOT sensitivity can be attained at this field – thus allowing, through phase biasing, optimal operation at any $B_a$ without blind spots.

Note that the first two terms of Eq. 3 can be rewritten as $I_2(B_a) = J_2(\varphi_2) - J_0$; namely, the trace of the maximum critical current $I_{c1}^{max}$ presented in bright color in Fig. 3b directly depicts the current-phase relation $J_2(\varphi_2)$ of junction 2. This correspondence holds for any form of current-phase relations[49] and therefore can provide a general tool, adding to existing techniques[52–55], for direct probing of current-phase relations in various unconventional superconductor systems.

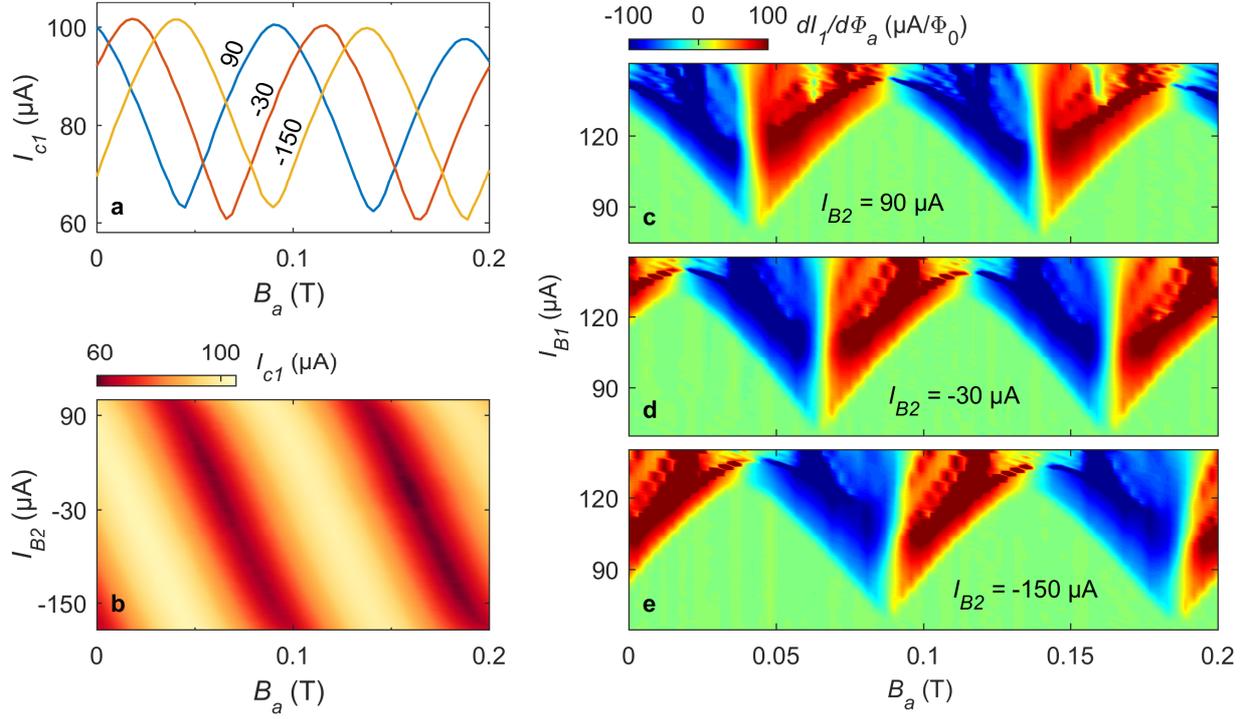

**Figure 3**. Electric tunability of the quantum interference pattern in mSOT device A. **(a)** Interference pattern $I_{c1}(B_a)$ at three values of the superconducting phase control current $I_{B2} = -150, -30,$ and $90$ µA at $I_{B4} = 0$. **(b)** Color coded plot of $I_{c1}(I_{B2}, B_a)$ showing a continuous shift of the interference pattern with the control current $I_{B2}$ at $I_{B4} = 0$. **(c-e)** The mSOT flux-to-current transfer function $dI_1/d\Phi_a$ maps for three values of $I_{B2}$ at $I_{B4} = 0$, showing the shift in field of the sensitive regions (dark red and blue) with varying the control current (see Movie M1).

In order to achieve a highly sensitive continuous operation with low flux noise $S_\Phi^{1/2} = S_I^{1/2}/|dI_1/d\Phi_a|$, one requires a high transfer function $|dI_1/d\Phi_a|$ and a low current noise $S_I^{1/2}$ at any value of the applied field. Figure 3c shows the measured color-coded $dI_1/d\Phi_a$ vs. bias current $I_{B1}$ and applied field $B_a$. The dark blue and red colors describe the regions of high transfer function, while the green color reflect regions of no sensitivity due to either $I_1 < I_{c1}(B_a)$ or the blind spots where $dI_1/d\Phi_a \approx 0$. By varying the control current $I_{B2}$, one can readily shift the sensitive regions to the desired value of $B_a$, as demonstrated in Figs. 3c-e. A continuous shift of the transfer function with $I_{B2}$ is presented in Movie M1.

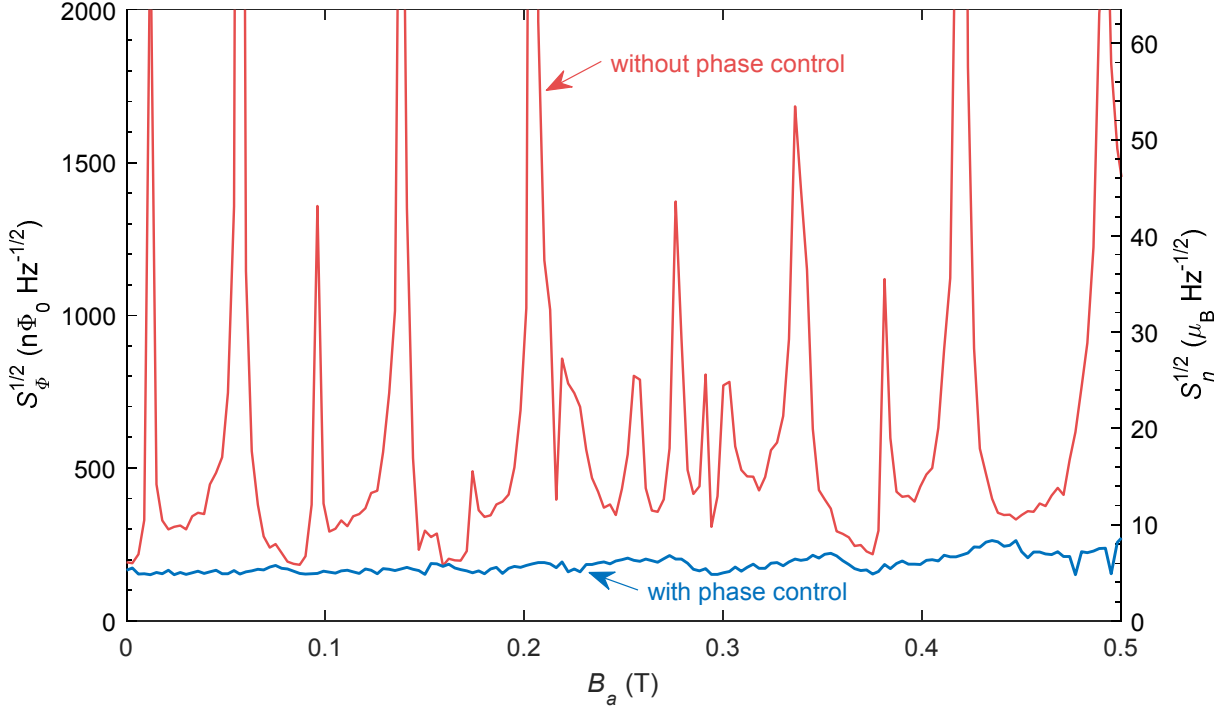

**Figure 4.** Flux noise (left) and spin noise (right) in mSOT device B. Red: The noise vs. applied field in the standard two-terminal configuration ($I_2 = I_4 = 0$) showing largely enhanced noise in the blind spots regions. Blue: Noise in the same device operating in the multi-terminal mode in which the superconducting phase control currents $I_2$ and $I_4$ are used to bias the mSOT to its optimal working point at each value of $B_a$. The blind spots are eliminated and a spin noise of ~ 5 to 8 $\mu_B/\text{Hz}^{1/2}$ is attained over the entire range of $B_a$ from 0 to 0.5 T, spanning six $\Phi_0$ periods.

A typical spectrum of the current noise $S_I^{1/2}$ of the mSOT is shown in Fig. 2c, displaying a $1/f$-like behavior at low frequencies $f \lesssim 100$ Hz followed by white noise at higher frequencies. By measuring the white noise level and the transfer function, the flux noise $S_\Phi^{1/2} = S_I^{1/2}/|dI_1/d\Phi_a|$ of the mSOT vs. $B_a$ is attained as shown in Fig. 4. The red curve displays the flux noise measured at $I_{B2} = I_{B4} = 0$, reflecting the conventional two-terminal SOT operation[24]. The large peaks in the noise show the blind spots regions of low transfer function. The corresponding spin noise[56] $S_n^{1/2} = S_\Phi^{1/2} r/r_e$ in units of $\mu_B/\text{Hz}^{1/2}$, assuming a point like magnetic dipole in the center of the SQUID loop, oriented perpendicular to it, is shown on the right axis, where $S_\Phi^{1/2}$ is in units of $\Phi_0/\text{Hz}^{1/2}$, $r_e = 2.82 \times 10^{-15}$ m is the classical electron radius, and $r = 89$ nm is the effective radius of the mSOT device B. The blue curve, in contrast, shows the performance of the mSOT in multi-terminal operation. For every $B_a$, the optimal values of the control currents $I_{B2}$ and $I_{B4}$ were attained in order to minimize the flux noise (see S6 and Movie M2 for details). The additional degrees of freedom provided by the control currents result in a remarkably uniform and low noise of 5 to 8 $\mu_B/Hz^{1/2}$ over an unprecedentedly wide continuous range of applied fields from 0 to 0.5 T spanning six $\Phi_0$ periods.

In summary, we have developed a novel multi-terminal/multi-junction SQUID-on-tip that provides highly sensitive continuous operation over a wide range of fields with spin sensitivity better than 8 $\mu_B/Hz^{1/2}$. Instead of the flux-bias used in FLLs in conventional SQUIDs, we introduce a superconducting phase control mode of operation that allows in-situ electric control of the quantum interference pattern without the need to modify the flux in the SQUID loop. As a result, nanoscale magnetic imaging can be performed with optimal sensitivity at any value of the local field without blind spots and without affecting the local field. By driving a current through a control terminal, the superconducting phase across one of the junctions is modified, thus shifting the quantum interference pattern. A single control terminal allows shifting the pattern by half a period while two terminals enable a shift by a full period. This control of the interference pattern through superconducting phase biasing opens the door for a new form of feedback control, the superconducting phase locked loop (SPLL). As a result, various additional noise reduction schemes analogous to the ones based on FLL protocols[39] can be implemented with mSOTs through an SPLL. The mSOT thus provides a new electrically tunable tool for highly sensitive quantitative imaging and study of magnetic phenomena on the nanoscale over an extremely large dynamic range.

**Supporting Information**.

This material is available free of charge via the Internet at http://pubs.acs.org.

mSOT fabrication; Estimate of mSOT inductance and noise limit; Electrical circuit; Movie M1 - interference pattern shift; Noise characterization; Movie M2 - optimal working point; Shift of the interference pattern by control current $I_4$; Shift of the interference pattern by concurrent application of $I_2$ and $I_4$.

**Corresponding Authors**

*Aviram Uri, E-mail: aviram.uri@weizmann.ac.il.

*Eli Zeldov, E-mail: eli.zeldov@weizmann.ac.il

**Present Addresses**

† Department of Physics, Columbia University, New York, New York 10027, USA.

ACKNOWLEDGMENTS

This work was supported by the US-Israel Binational Science Foundation (BSF grant 2014155), by the European Research Council (ERC) under the European Union's Horizon 2020 program (grant No 655416), and by Rosa and Emilio Segré Research Award.

# Supporting Information

# Electronically tunable multi-terminal SQUID-on-tip

*Aviram Uri[1]\*, Alexander Y. Meltzer[1], Yonathan Anahory[1], Lior Embon[1,†], Ella O. Lachman[1], Dorri Halbertal[1], Naren HR [1], Yuri Myasoedov[1], Martin E. Huber[1,2], Andrea Young[1,3], and Eli Zeldov[1]\**

[1]Department of Condensed Matter Physics, Weizmann Institute of Science, Rehovot 7610001, Israel

[2]Departments of Physics and Electrical Engineering, University of Colorado Denver, Denver, Colorado 80217, USA

[3]Department of Physics, Broida Hall, University of California, Santa Barbara, CA 93106-9530

**S1.    mSOT fabrication**

A 1.0 mm outer diameter, 0.4 mm inner diameter quartz tube with four $0.1 \times 0.15$ mm$^2$ grooves equally spaced on its circumference (see Fig. 1b) was laser heated and pulled to a diameter ranging from 100 to 200 nm using a Sutter Instruments P-2000 puller. A thin film of Pb was deposited on the pulled pipette in a three-step self-aligned thermal evaporation process (15 to 16 nm thickness for the leads and 12 to 14 nm for the apex ring). The pipette was cooled to cryogenic temperatures during the evaporation using a flowing He cryostat to ensure good adhesion of the Pb to the quartz surface[1]. In this work we present three devices. Device A had a field period of 97 mT corresponding to an effective diameter of 165 nm and a critical current of 105 µA. Device B had a field period of 84 mT corresponding to an effective diameter of 177 nm and a critical current of 61 μA. Device C had a field period of 114 mT, effective diameter of 152 nm and a critical current of 69 μA.

**S2.  Estimate of mSOT inductance and noise limit**

To estimate the geometric inductance $L_g$ of our devices we approximate the mSOT as a loop of wire with circular cross section, giving rise to a very low $L_g = \mu_0 R \left[\log\left(\frac{8R}{r}\right) - 2\right] \approx 0.2$ pH, where $R \approx 82$ nm is the loop radius, and $r \approx 13$ nm is the radius of the wire. The largest possible circulating current in the loop is $J_0 \sim I_c/2 \approx 50$ μA, resulting in the upper bound of self-induced flux in the loop of $\Phi = L_g J_0 \approx 10^{-17}$ Wb $\approx 5 \times 10^{-3}$ $\Phi_0$. The contribution of the geometric inductance to the shift of the interference pattern is therefore negligible.

Similar to the two-junction SOTs[1,2], the total inductance $L = L_g + L_k$ of the mSOT is governed by the kinetic inductance $L_k$. The inductance $L$ can be estimated from the modulation depth of the critical current with the magnetic field, which for the case of mSOT device A was ~39%. For an ideal four-junction SQUID this modulation depth corresponds[3] to $\beta_L = \frac{I_c L}{\Phi_0} \approx 0.4$ resulting in $L \approx$ 8.5 pH. Since similarly to SOTs the mSOT has very low inductance and capacitance[1], its ultimate flux noise is expected to be limited by shot noise $S_\Phi^{1/2} = \sqrt{hL}$. For the above value of $L$ this results in $S_\Phi^{1/2} = 36$ n$\Phi_0$Hz$^{-1/2}$ which is about 5 times lower than the measured mSOT flux noise in Fig. 2c.

**S3.  Electrical circuit**

Figure 2a shows the schematic electrical circuit used for the bias and the readout of the mSOT. The current bias $I_{B1}$ (voltage source followed by a series resistor of 5 kΩ (device A) or 2.1 kΩ (devices B and C)) was applied to the mSOT in parallel with the shunt resistor $R_{s1} = 5\,\Omega$, which is much smaller than the normal state resistance of the mSOT junctions typically ranging from 50 to 100 Ω. Terminal 3 was grounded, thus acting as a drain. This gave a steep load line, close to voltage bias, allowing non-hysteretic operation of the underdamped mSOT which exhibits negative differential resistance[1] (see Fig. 2b). The current $I_1$ was measured using a SQUID series array amplifier (SSAA)[4]. When the mSOT is in the superconducting state, most of the bias current $I_{B1}$ flows through the device and only a small part of it is diverted to the shunt resistor $R_{s1}$. The parasitic resistance $R_p$ in series with the device and its ratio to the shunt resistance $R_{s1}$ determines the amount of current diverted to the shunt. This is reflected in the linear part of $I_1(I_{B1})$ shown in Fig. 2b. When the mSOT is biased above the critical current and into the voltage state, the device becomes resistive and a larger portion of the bias current $I_{B1}$ is diverted to the shunt resistor $R_{s1}$.

The superconducting phase control currents $I_2$ and $I_4$ were applied as shown schematically in Fig. 2a. In case of device A, the shunt resistor $R_{s2} = 5\,\Omega$ was used and $R_{s4}$ was absent. For devices B and C, the control bias currents $I_{B2}$ and $I_{B4}$ were applied directly, with no $R_{s2}$ and $R_{s4}$ resistors.

**S4. Movie of the interference pattern shift**

Movie M1 shows the device A transfer function $dI_1/d\Phi_a$ maps for increasing values of $I_{B2}$, at $I_{B4} = 0$, showing the shift in field of the mSOT interference pattern as the control current is varied. Figures 3e-h show four frames from this movie.

## S5. Noise characterization

Noise characterization of the mSOT can be very time consuming if the full noise spectrum is acquired at every point in the large parameter space ($I_{B1}, I_{B2}, I_{B4}, B_a$). To overcome this, white noise in a frequency band centered at 10 kHz was measured using a diode as a rectifier, as shown in Fig. S1, allowing for a fast measurement of the noise (~1 ms per data point). The SSAA output signal was fed into the circuit input $V_{in}$, band-pass filtered around 10 kHz, amplified by $G_1 = 2 \cdot 10^4$ V/V, high-pass filtered to remove offsets and rectified using a diode (Herotek DZM040AA for device A and PMEG6010 Schottky diode for devices B and C). The rectified signal was amplified by $G_2 = 10$ V/V and averaged using a low-pass filter. The output DC signal $V_{out}$ was proportional to the device noise at 10 kHz and the proportionality factor was determined by a calibration measurement against the full spectral density like the one shown in Fig. 2c. Figure S2 shows electrical characteristics of mSOT device A over the full range of applied field with corresponding measurement of the white current noise in Fig. S2c.

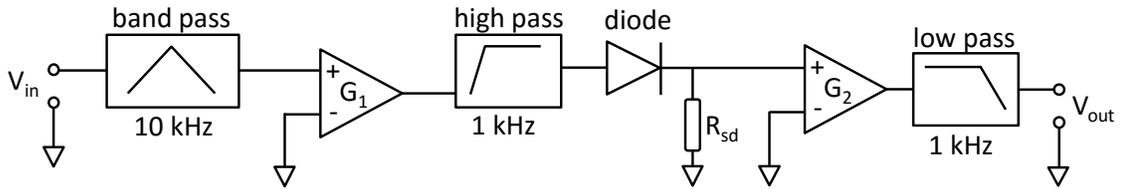

**Figure S1.** Electrical circuit used to measure white noise using a diode rectifier.

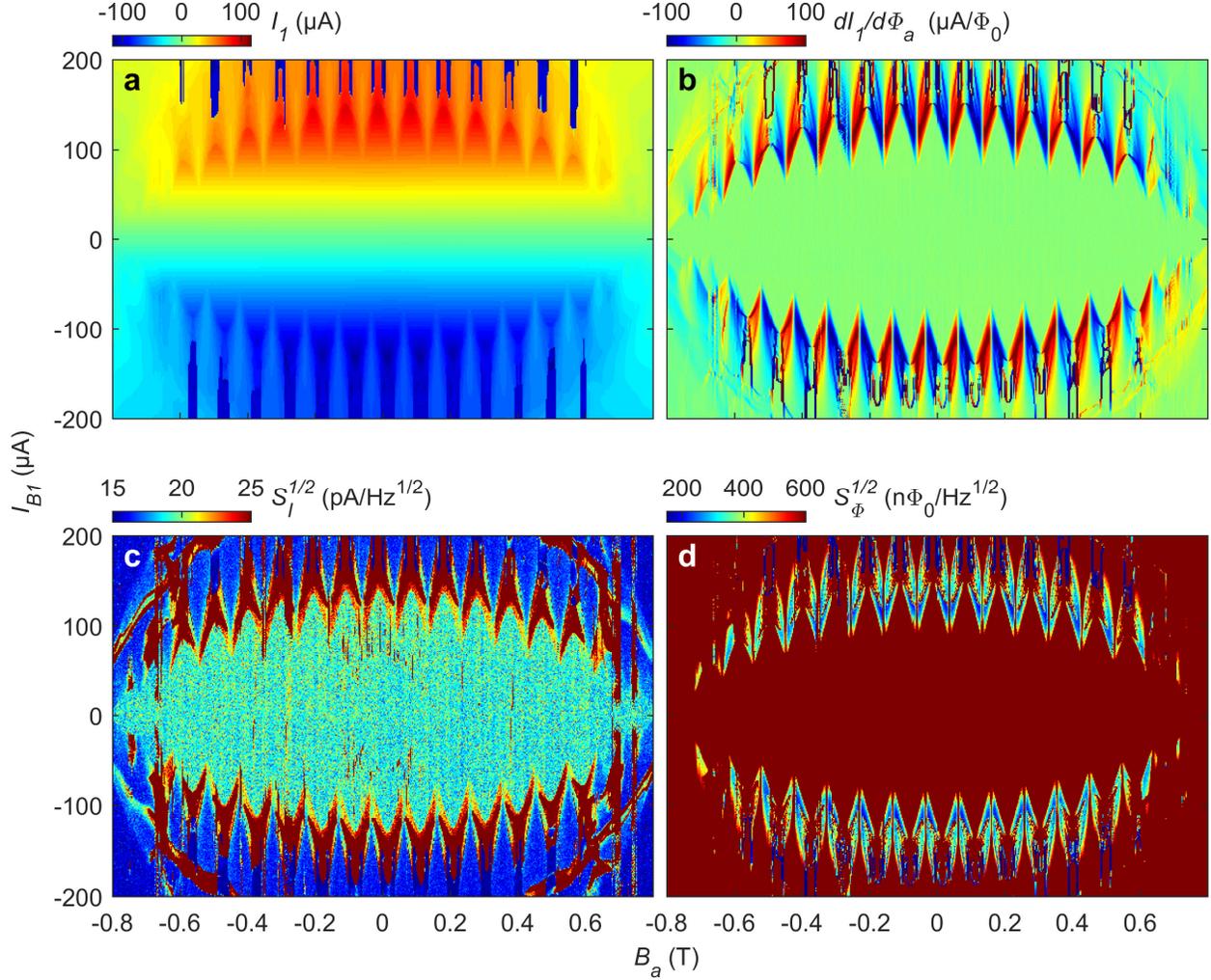

**Figure S2.** Electrical characteristics of mSOT device A as function of applied field $B_a$ and current bias $I_{B1}$, keeping $I_{B2} = I_{B4} = 0$. **(a)** The mSOT current $I_1$. **(b)** The transfer function $dI_1/d\Phi_a$ derived numerically from (a). **(c)** Current noise $S_I^{1/2}$ measured using the scheme in Fig. S1. **(d)** Flux noise $S_\Phi^{1/2} = S_I^{1/2}/|dI_1/d\Phi_a|$ attained numerically from (b) and (c).

### S6. Optimal working points

For every field value $B_a$, we define the optimal working point as the set of bias and control currents $I_{B1}$, $I_{B2}$ and $I_{B4}$ that result in the minimal flux noise $S_\Phi^{1/2}(I_{B1}, I_{B2}, I_{B4}; \Phi_a)$. Movie M2 presents this concept applied to device A, keeping $I_{B4} = 0$ for simplicity. Several frames from this movie are given in Figs. S3e-h, showing color-coded flux noise $S_\Phi^{1/2}(I_{B1}, I_{B2})$ at various applied fields $B_a$ with corresponding transfer function $dI_1/d\Phi_a(I_{B1}, I_{B2})$ shown in Figs. S3a-d. The central region in the figures with $dI_1/d\Phi_a = 0$ (green) represents the superconducting state of the mSOT with $I_1 < I_{c1}(I_{B2}, I_{B4}, \Phi_a)$, as described theoretically in Ref. 3. The optimal working point with lowest flux noise is denoted by a white circle. . Figure 4 shows the resulting flux noise along the optimal working point trajectory in device B.

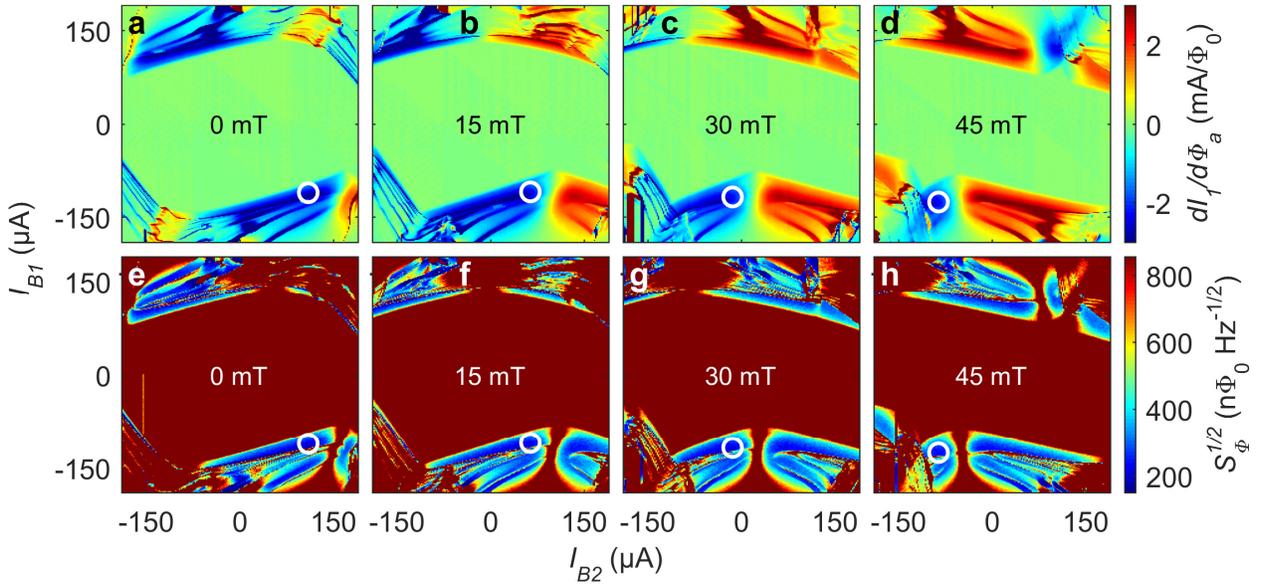

**Figure S3.** Color coded transfer function $dI_1/d\Phi_a(I_{B1}, I_{B2})$ (**a-d**) and flux noise $S_\Phi^{1/2}(I_{B1}, I_{B2})$ (**e-h**) at different applied fields $B_a$ as indicated on the plots (device A). For each applied field $B_a$, the optimal working point with the lowest flux noise is marked by a white circle (frames taken from Movie M2).

### S7. Shift of the interference pattern by control current $I_4$

The superconducting phase control current $I_4$ shifts the interference pattern similarly to $I_2$ but in the opposite direction, as demonstrated in Fig. S4a, which shows the behavior in device C over an extended range of $I_4$. Since device C was connected without shunt resistors $R_{s2}$ and $R_{s4}$, the control currents $I_4 = I_{B4}$ and $I_2 = I_{B2}$. The dashed curve in Fig. S4a, $B_a^{max}(I_4)$, traces the field location of the maximum of the critical current $I_{c1}^{max}$ as the interference pattern is shifted by $I_4$. The curve $B_a^{max}(I_4)$ thus describes the field shift of the entire interference pattern as induced by the phase bias. As derived in Ref. 3 and as seen in Fig. S4, $B_a^{max}(I_4)$ has a cusp and changes the shift direction at the demarcation line $I_4 = J_{03} - J_{04} = -6$ µA, where $J_{03}$ and $J_{04}$ are the critical currents of junctions 3 and 4 respectively. It can be readily shown[3] that for $I_4 < J_{03} - J_{04}$ the $I_{c1}^{max}$ along the $B_a^{max}(I_4)$ curve is constant, $J_4 = -J_{04}$, $\varphi_4 = -\pi/2$, and $\varphi_3$ is given by $J_{03} \sin \varphi_3 = -J_{04} - I_4$. In this region, the $B_a^{max}(I_4)$ curve represents the current-phase relation of junction 3. An equivalent case of using the control terminal 2 for shifting the interference pattern in the regime of $I_2 < J_{02} - J_{01}$ is presented in Fig. 3b.

Above the demarcation line $I_4 > J_{03} - J_{04}$ the $I_{c1}^{max}$ along the $B_a^{max}(I_4)$ decreases linearly with $I_4$, $J_3 = -J_{03}$, $\varphi_3 = -\pi/2$, and $\varphi_4$ is given by[3] $J_{04} \sin \varphi_4 = I_4 - J_{03}$. In this region the $B_a^{max}(I_4)$ curve reflects the current-phase relation of junction 4. Since $I_{c1}^{max}$ is not constant, this regime is less convenient for practical applications.

### S8. Shift of the interference pattern by concurrent application of $I_2$ and $I_4$

By utilizing the two control currents $I_2$ and $I_4$ concurrently, an extensive control of the interference pattern can be attained. In particular, a single control current allows a shift of the interference pattern by about half a period. By using both control currents, a shift of a full period can be attained[3]. This is demonstrated in Fig. S4b, where $B_a^{max}(I_2, I_4)$ is plotted in color code. The vertical black dashed line shows the values of $B_a^{max}(I_4)$ along the dashed line in Fig. S4a for $I_2 = 0$. The maximum variation of $B_a^{max}(I_2, I_4)$ along any vertical or horizontal line in Fig. S4b corresponds to about half a period. However, by utilizing both control currents along a diagonal curve like the dotted white line, a shift of $B_a^{max}$ of about a full period is achieved. In addition to attaining optimal sensitivity at any value of the applied field, this wide control opens the possibility of implementation of noise reduction schemes based on the described SPLL similar to the common FLL methods[5].

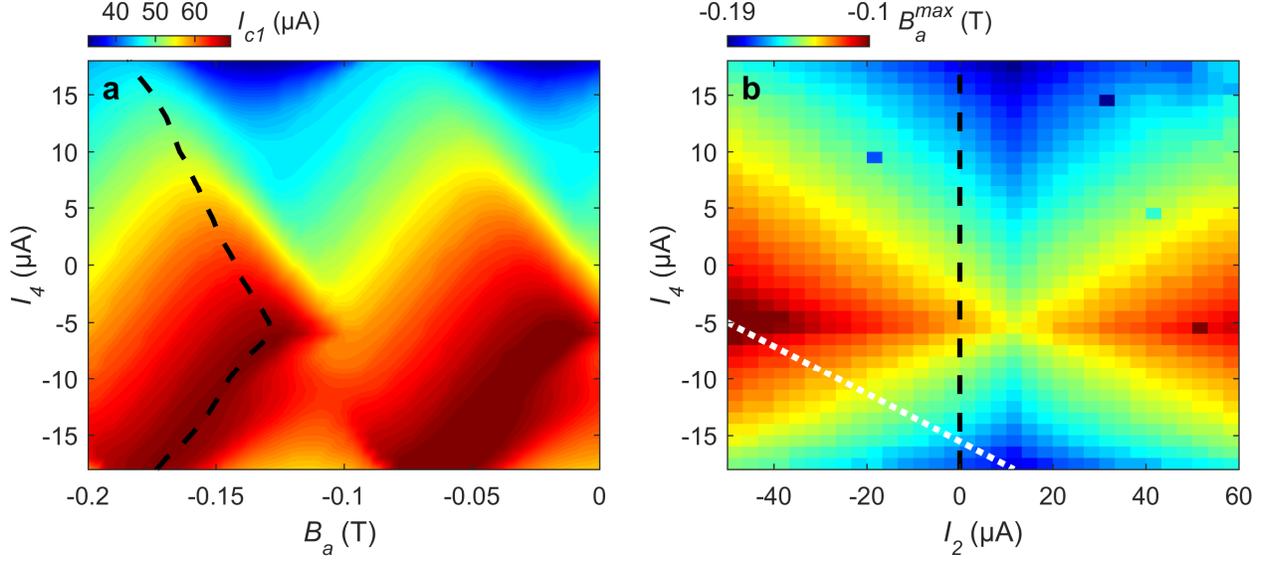

**Figure S4.** The interference pattern shift in device C. **(a)** Color coded $I_{c1}(B_a, I_4)$ at $I_2 = 0$ showing positive and negative field shifts of the interference pattern with $I_4$ for $I_4 \lesssim -6\,\mu A$ and $I_4 \gtrsim -6\,\mu A$ respectively. The dashed line, $B_a^{max}(I_4)$, traces the field location of the maximum of the critical current $I_{c1}^{max}$. **(b)** Color coded $B_a^{max}(I_2, I_4)$ showing the extent of the shift of the interference patterns upon application of the two superconducting phase control currents. The vertical dashed line shows the values of $B_a^{max}(I_4)$ along the dashed line in (a) for $I_2 = 0$. The dotted diagonal line shows the trajectory of the concurrent use of the two superconducting phase control currents along which a shift of the interference pattern of about a full period is attained.

**Present Addresses**

[†] Department of Physics, Columbia University, New York, New York 10027, USA.